\documentclass[aps,prd,twocolumn,superscriptaddress,showpacs]{revtex4-1}

\usepackage{amsmath}
\usepackage{graphicx}
\usepackage{epstopdf}
\usepackage{color}
\usepackage{slashed}
\usepackage{CJK}

\definecolor{purple}{rgb}{0.4,0,0.6}
\definecolor{blue}{rgb}{0.0,0,1.0}
\definecolor{prdblue}{rgb}{0.133,0.118,0.498}

\usepackage[colorlinks=true, pdfstartview=FitV, linkcolor=prdblue, citecolor= prdblue, urlcolor=prdblue]{hyperref}

\begin{document}

\begin{CJK*}{GBK}{song}

\title{New Insight  about the Effective Restoration of  $U_{A}(1)$ Symmetry}

\author{Xiang Li}
\affiliation{Department of Physics and State Key Laboratory of
Nuclear Physics and Technology, Peking University, Beijing 100871,
China}
\affiliation{Collaborative Innovation Center of Quantum Matter, Beijing 100871, China}
\author{Wei-jie Fu}
\affiliation{School of Physics, Dalian University of Technology, Dalian 116024, China}
\author{Yu-xin Liu}
\email[Corresponding author: ]{yxliu@pku.edu.cn}
\affiliation{Department of Physics and State Key Laboratory of
Nuclear Physics and Technology, Peking University, Beijing 100871,
China}
\affiliation{Collaborative Innovation Center of Quantum Matter, Beijing 100871, China}
\affiliation{Center for High Energy Physics, Peking University, Beijing 100871, China}

\date{\today}

\begin{abstract}
The effective restoration of the $U_A(1)$ symmetry is revisited by implementing the functional renormalization group approach combining with the 2+1 flavor Polyakov-loop quark-meson model.
A temperature-dependent 't Hooft term is taken to imitate the restoration of the $U_{A}(1)$ symmetry.
Order parameters, meson spectrum and mixing angles, especially the pressure and the entropy density of the system
are calculated to explore the effects of different $U_{A}(1)$ symmetry restoration patterns.
We show then that the temperature for the restoration of the $U_{A}(1)$ symmetry
is much higher than that for the  chiral symmetry $SU_{A}(3)$.
\end{abstract}

%

\maketitle

\section{Introduction}

Studies on the strongly interacting system (QCD) have been attractive over decades,
since a full understanding of QCD system is crucial for exploring the fundamental structure of nature.
Due to the non-perturbative character and special vacuum structure of QCD, many problems of QCD system remain unsettled,
for instance, $U_{A}(1)$ anomaly and its restoration is a long-standing one in this area~\cite{Hooft:1976PRL,Witten:1979NP,Gross:1981RMP,Schafer:1998Review,Pisarski:1984PRD,Brown:1990PRL,Khazeev:2010AP}.

In QCD system, the spontaneous breaking of chiral symmetry $SU_{A}(3)$ leads to eight pseudo-Goldstone bosons.
The left axial symmetry  $U_{A}(1)$ is violated by quantum anomaly and results in a heavy meson, i.e., $\eta'$~\cite{Witten:1979NP}.
However, it's predicted in Ref.~\cite{Gross:1981RMP} that $U_{A}(1)$ symmetry can be effectively restored at high temperature due to the suppression of the instanton density of the QCD vacuum.
This prediction is proved later in many Lattice QCD simulation results~\cite{Alles:1997NP,Bhattacharya:2014PRL,Aoki:2008PL,Brandt:2016JHEP,Suzuki:2018POS}, whereas the specific temperature for $U_{A}(1)$ to be restored is still far from clear and requires more investigations.

Depending on whether $U_{A}(1)$ symmetry is restored before the chiral phase transition, the whole system will belong to different universal class and the order of the chiral phase transition can be changed, and then leads to different Columbia plots~\cite{Pisarski:1984PRD,Brown:1990PRL}.
Various quantities (such as the topological susceptibility and the mesonic correlators~\cite{Alles:1997NP,Bhattacharya:2014PRL,Aoki:2008PL,Brandt:2016JHEP,Suzuki:2018POS}) have been calculated in Lattice QCD to investigate the restoration of the $U_{A}(1)$ symmetry. The calculated topological susceptibility and the mass splitting between the scalar and pseudo-scalar mesons (for example, $a_{0}$ and $\pi$) all tend to decrease near the chiral phase transition. These results indicate a partial restoration of the $U_{A}(1)$ symmetry near the chiral phase transition, but more numerical efforts are needed to reach a definite conclusion.

Besides Lattice QCD method, continuum field approaches such as the Dyson-Schwinger equation approach  and the functional renormalization group approach have also been taken to survey the $U_{A}(1)$ problems (see, e.g., Refs.~\cite{Pawlowski:1998PRD,Benic:2011PRD,Alkofer:2008EPJ,Horvatic:2019PRD}). Compared with Lattice QCD method, continuum field approach usually requires less numerical efforts and the chiral symmetry can be implemented easily.

In addition to the first principle approach mentioned above, $U_A(1)$ symmetry has also been investigated via effective models (see, e.g., Refs.~\cite{Bielich:2000PRL,Costa:2004PRD,Schaefer:2009PRD,Fukushima:2001PRC,Chen:2009PRD,Nagahiro:2005PRL,Nagahiro:2006PRC,Sakai:2013PRC,Fu:2008PRD,Ciminale:2008PRD,Gupta:2010PRD,Fukushima:2008PRD,Lenaghan:2000PRD,Nicola:2016JHEP,Fabian:2017PRD,Dunne:2010PRD,Guo:2012PLB,Mitter:2014PRD}). Simple phenomenological models (for example, linear sigma model, NJL model and quark-meson model) are taken to approximate the QCD system and the $U_{A}(1)$ anomaly is usually implemented via the 't Hooft term~\cite{Hooft:1976PRL}. It is shown that order parameter and meson spectrum will be significantly affected when the effective restoration of the $U_{A}(1)$ symmetry is considered, and several efficient signals have been predicted in heavy ion collision experiments to detect the $U_{A}(1)$ restoration effect (see, e.g., Ref.~\cite{Bielich:2000PRL}). And it has also been shown in Ref.~\cite{Fukushima:2001PRC} that the $U_{A}(1)$ symmetry remains broken when the chiral transition happens, but this prediction is somehow model-dependent and more detailed investigation is still needed.

In this work, we employ the functional renormalization group (FRG) approach~\cite{Wetterich:1993Review,Litim:2000PLB,Litim:2006JHEP,Litim:2001PRD,Litim:2001JHEP,Litim:2011PRD,Stokic:2010EPJ,Jan:2007Review,Gies:2012Review} combining with the 2+1 flavor Polyakov-loop quark-meson model~\cite{Tawfik:2014PRC,Herbst:2014PLB,Fabian:2017PRD,Fu:2018,Fu:20182} (PQM) to calculate the order parameters, the meson spectrum and mixing angles, the pressure and entropy density of the system with a restoring $ U_A(1)$ symmetry.
It's well known that the PQM model is an effective approximation of QCD at low energy region
and the FRG approach can go beyond the widely used mean-field approximation.
The effective restoration of the $U_{A}(1)$ symmetry is imitated via a temperature-dependent 't Hooft term deduced from Lattice QCD simulations and theoretical derivations~\cite{Bielich:2000PRL,Costa:2004PRD,Rai:2018arxiv}. Compared with previous works, the restoration of the $U_{A}(1)$ symmetry is viewed from a new perspective: the pressure and the entropy density of the system are taken to identify the effect of the $U_{A}(1)$ symmetry restoration.
As we will see, unphysical thermodynamical result will appear if the $U_{A}(1)$ symmetry is restored before the chiral phase transition. We show then that the $U_{A}(1)$ symmetry is still broken as the chiral phase transition happens.

The remainder of this paper is organized as follows. In Sec.~\ref{Sec:theory}, we introduce briefly the main aspects of the FRG approach and the PQM model. Some discussions about the temperature-dependent 't Hooft term are also given. In Sec.~\ref{Sec:res}, we show the obtained results and discuss the underlying mechanism. In Sec.~\ref{Sec:sum}, we give our summary and some remarks.

\section{Theoretical Framework}\label{Sec:theory}

In this section, we describe concisely the FRG approach and the PQM model for self-consistency, and more details can be found in Refs.~\cite{Fabian:2017PRD,Fu:2018,Fu:20182,Kamikado:2015JHEP,Xiang:2019PRD}.
Briefly speaking, the PQM model can be recognized as a linear sigma model coupled with a static gluon background field, with the gluon field being integrated out. Its Lagrangian reads simply
\begin{eqnarray}\label{Eq:lag}
\mathcal{L}=\ &&\bar\psi(\slashed \partial - i\gamma_{4} A_{4} + g\Sigma_{5})\psi
+ {\rm Tr}[\mathcal{\partial}_\mu\Sigma\cdot (\mathcal{\partial}_{\mu} \Sigma)^{\dagger}]  \nonumber  \\
& & +\ U(\Sigma) - h_{x} \sigma_{x} - h_{y} \sigma_{y} - c_{a} \xi +V_{poly}(\Phi)\, ,
\end{eqnarray}
where $\psi$ represents the 3 flavor (u,d,s) quark field, and $\Sigma=(\sigma_{a} + i\pi_{a})T^{a}$ is a matrix contains scalar and pseudo-scalar meson nonets. $V_{poly}$ is a phenomenological potential of Polyakov-loop $\Phi$ which is added to imitate the confinement effect. For specific definition of other terms in Eq.~(\ref{Eq:lag}), see Refs.~\cite{Kamikado:2015JHEP,Xiang:2019PRD}.

Among all the terms in Eq.~(\ref{Eq:lag}), the 't Hooft term $c_{a}\xi$ is quite special:
$\xi={\rm det}(\Sigma)+{\rm det}(\Sigma^\dagger)$ breaks the $U_{A}(1)$ symmetry explicitly
and the $c_{a}$ is a parameter measuring the strength of the axial anomaly.
This term origins from the non-trivial vacuum structure of the gauge field which can be characterized by the so called winding numbers. The topologically different vacuums can be linked by instanton which is also the gaussian stable point of the path integral and should contribute to the partition function.
However, the $U_{A}(1)$ charge of quarks is not conserved in an instanton background and thus leads to the effective $U_A(1)$ breaking term $c_{a}\xi$~\cite{Hooft:1976PRL,Schafer:1998Review}.

When the temperature effect is considered, the contribution from the instantons will get suppressed by Debye screening, and the $U_{A}(1)$ symmetry can be effectively restored~\cite{Gross:1981RMP}.
In order to take the restoration of the $U_{A}(1)$ symmetry into consideration,
it is usual to parameterize the $c_{a}$ as a function of temperature~\cite{Bielich:2000PRL,Costa:2004PRD,Rai:2018arxiv}.
Lattice QCD simulations show that the topological susceptibility is nearly unchanged at low temperature~\cite{Alles:1997NP}, and theoretical derivations predict a exponentially decay of the instanton density at high temperature~\cite{Gross:1981RMP}.
Combining these two aspects, we employ the form proposed in Ref.~\cite{Rai:2018arxiv} which reads
\begin{eqnarray}\label{Eq:anomal}
c_{a}(T)=\begin{cases}    c_{a}(0)\, ,  & \mbox{$T < T_{\textrm{r}}$} \, ; \\
c_a(0) \, {\rm exp}\big[-\frac{(T - T_{\textrm{r}})^{2}}{b^{2}}\big] \, ,\quad  & \mbox{$T > T_{\textrm{r}}$} \, ;
 \end{cases}\,
\end{eqnarray}
where $c_{a}(0)$ is a constant obtained by fitting meson spectrum at vacuum.
$T_{\textrm{r}}$ and $b$ are two free parameters: $T_{\textrm{r}}$ is simply the starting temperature for the $U_{A}(1)$ symmetry to be restored and $b$ determines the restoration speed.
In this work, we will mainly tune the $T_{\textrm{r}}$ to control the restoration pattern of the $U_{A}(1)$ symmetry.
Note that there exist other ways to parameterize the $c_{a}$ (see, e.g., Refs.~\cite{Bielich:2000PRL,Costa:2004PRD}),
but they are all similar with each other and would not make much difference to the results.

It is worthy mentioning that the $c_{a}$ will receive contributions from thermal fluctuations and then grow a temperature dependence even without instanton effect (see, e.g.,  Refs.~\cite{Fejos:2016PRD,Jiang:2012PRD}).
So Eq.~(\ref{Eq:anomal}) can be seen as a crude approximation of $U_{A}(1)$ restoration which considers instanton effect only and neglects thermal fluctuation contributions.

After introducing the Lagrangian, the FRG approach can be employed to study the thermodynamics of the PQM system.
A functional evolution equation for the effective action $\Gamma_{k}$ is derived to integrate different momentum shell out gradually~\cite{ Wetterich:1993Review}, which reads
\begin{eqnarray}\label{Eq:frg}
\partial_{k} \Gamma_{k} =\frac{1}{2}{\rm Tr}\Big{[}\frac{\partial_{k} R_{k}^{B}}{\Gamma^{2B}_{k} + R_{k}^{B}}\Big{]}
-{\rm Tr}\Big{[}\frac{\partial_{k} R_{k}^{F}}{\Gamma^{2F}_{k} + R_{k}^{F}}\Big{]}\, ,
\label{Eq:wet}
\end{eqnarray}
where $R_{k}^{B,F}$ are the momentum-dependent mass terms assigned to the quarks and mesons, and $\Gamma_{k}^{2B},\Gamma_{k}^{2F}$ denote the second derivatives of $\Gamma_{k}$ with respect to the corresponding fields.
Compared with the traditional mean-field approximation, FRG approach can incorporate mesons fluctuations into the evolution.
It's well known that mesons such as pion dominate at low temperature and their fluctuations will affect the system significantly (see, e.g., Refs.~\cite{Kamikado:2015JHEP,Xiang:2019PRD}). Thus the FRG approach is usually known to be a method beyond mean-field approximation.

It is usually impossible to solve Eq.~(\ref{Eq:wet}) exactly, we take then the local potential approximation (LPA) in this paper to simplify the problem. The truncated $\Gamma_{k}$ reads
\begin{eqnarray}\label{Eq:eff}
\Gamma_{k} = \int d^{4}x &&\ \bar\psi(\slashed \partial - i\gamma_{4} A_{4} + g\Sigma_{5})\psi
+\ {\rm Tr}[\mathcal{\partial}_{\mu} \Sigma \cdot (\mathcal{\partial}_{\mu} \Sigma)^{\dagger}]   \nonumber   \\
&&+\ U_{k}(\Sigma) - h_{x} \sigma_{x} - h_{y} \sigma_{y} - c_{a} \xi + V_{poly}(\Phi)\, ,        \nonumber   \\
\end{eqnarray}
where $\sigma_{x}$ and $\sigma_{y}$ are related to the meson fields via a rotation
\begin{eqnarray}
\begin{pmatrix}
\sigma_{x}\\
\sigma_{y}
\end{pmatrix}
=\frac{1}{\sqrt{3}}
\begin{pmatrix}
1& \sqrt{2}\\
-\sqrt{2} & 1
\end{pmatrix}
\begin{pmatrix}
\sigma_{8} \\
\sigma_{0}
\end{pmatrix}\, .
\end{eqnarray}
The only thing flows with scale $k$ in Eq.~(\ref{Eq:eff}) is the $U_{k}(\Sigma)$ term.
Substituting Eq.~(\ref{Eq:eff}) into Eq.~(\ref{Eq:frg}) one obtains the flow equation for $U_{k}$ as
\begin{eqnarray}\label{Eq:effpotflow}
\partial_{k}U_{k} = &&\ \frac{k^4}{12\pi^2}\Big{\{}\sum_{b} \frac{1}{E_{b}}[1 + 2n_{b} (E_{b})]        \nonumber \\
&&\,\,\, -\sum_{f=u,d,s}\frac{4N_{c}}{E_{f}}[1 - 2\tilde{n}_{f}(E_{f},\Phi)]  \Big{\}}\, .  \qquad
\end{eqnarray}
Note that $U_{k}(\Sigma)$ will develop a dependence on $\Phi$ via quark's fluctuations in the last line of Eq.~(\ref{Eq:effpotflow}).
To accomplish the calculation, we adopt the 3-dimensional infrared regulators proposed in Refs.~\cite{Litim:2000PLB,Litim:2006JHEP,Litim:2001PRD,Litim:2001JHEP,Litim:2011PRD,Stokic:2010EPJ}.
The equation can then be solved numerically by Taylor method~\cite{Fabian:2017PRD,Mitter:2014PRD,Fu:2018,Fu:20182,Kamikado:2015JHEP,Xiang:2019PRD} and the parameters used in this work are the same as those in Ref.~\cite{Xiang:2019PRD}.
We can then get the thermodynamic property of the PQM system after the full $U_{0}(\Sigma,\Phi)$ is obtained.

\section{Result}\label{Sec:res}
After Eq.~(\ref{Eq:effpotflow}) is solved, various quantities can be obtained via the effective potential $\tilde{U}(\sigma_{x},\sigma_{y},\Phi)$, which reads
\begin{eqnarray}\label{Eq:effpot}
\tilde{U}(\sigma_{x}, \sigma_{y}, \Phi)=\ &&U_{0}(\sigma_{x},\sigma_{y},\Phi) + V_{poly}(\Phi)    \nonumber   \\
&&-\, h_{x} \sigma_{x} - h_{y} \sigma_{y} - c_{a} \frac{\sigma_{x}^{2} \sigma_{y}}{2\sqrt{2}}\, .
\end{eqnarray}
Note that the 't Hooft term $c_{a}\xi$ has been reduced to the last term in Eq.~(\ref{Eq:effpot})
since only the $\sigma_{x}$ and $\sigma_{y}$ remain nonzero now.
The quantities $\tilde{\sigma}_{x}$ and $\tilde{\Phi}$ corresponding to the minimums of the effective potential $\tilde{U}$ are usually taken as the order parameters for the chiral and the deconfinement phase transition respectively. And the chiral pseudo-critical temperature $T_{c}^{\chi}$ extracted from the inflection point of $\tilde{\sigma}_{x}$ is
$T_{c}^{\chi}=208\ {\rm MeV}$ if the anomaly strength $c_{a}$ keeps constant.

In order to investigate the effects of different $U_{A}(1)$ symmetry restoration patterns,
we set the $U_A(1)$ restoration temperature $T_{\textrm{r}}$ to three typical values $150, 200$
and $250\ {\rm MeV}$. Another parameter $b$ is set to $50$ MeV and different choices of the $b$
would not induce much difference.

\subsection{ORDER PARAMETERS AND MESON SPECTRUM}\label{Sec:order}

\begin{figure}[htb]
\centering
\includegraphics[width=0.45\textwidth]{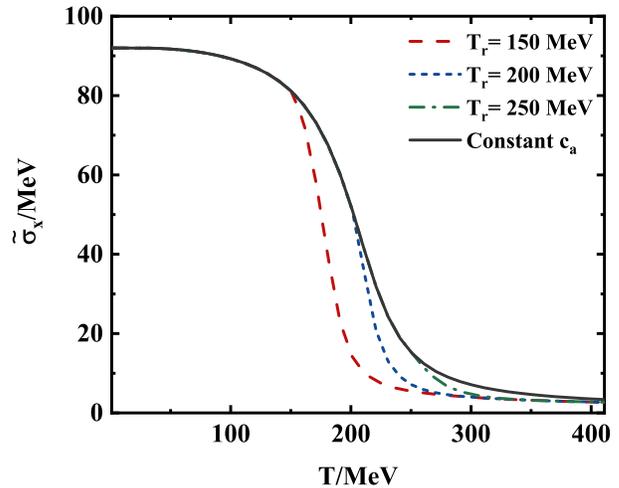}
\vspace*{-3mm}
\caption{(color online) Calculated $\tilde{\sigma}_{x}$ as functions of temperature at several values of $T_{\textrm{r}}$.}\label{order}
\end{figure}

\begin{figure}[htb]
\centering
\includegraphics[width=0.45\textwidth]{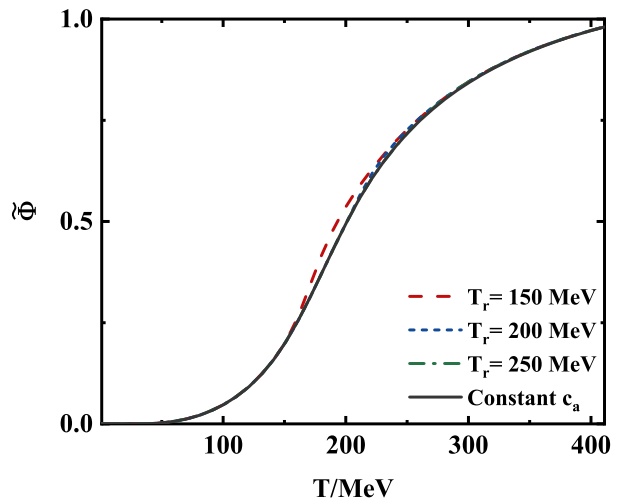}
\vspace*{-3mm}
\caption{(color online) Calculated Polyakov loop $\tilde{\Phi}$ as functions of temperature at several values of $T_{\textrm{r}}$.}\label{poly}
\end{figure}

The calculated order parameters are displayed in Figs.~\ref{order} and \ref{poly}.
As we see directly from the Fig.~\ref{order}, the $\tilde{\sigma}_{x}$  decreases monotonously with the rising of temperature.
This simply means that the chiral symmetry $SU_{A}(3)$ is getting restored gradually.
After the restoration of $U_{A}(1)$ symmetry is considered,
$\tilde{\sigma}_{x}$ will get reduced significantly compared with the constant $c_{a}$ case  and the chiral pseudo-critical temperature is then shifted to a lower value:  the $T_c^\chi$ for $T_{\textrm{r}} =150\ {\rm MeV}$ case is lowered to
$177\ {\rm MeV}$, while the $T_{c}^{\chi}$ for $T_{\textrm{r}} = {200,250\ {\rm MeV}}$ cases is nearly unchanged compared with the constant $c_{a}$ case.
These effects of $U_{A}(1)$ restoration are also predicted in Refs.~\cite{Bielich:2000PRL,Costa:2004PRD,Rai:2018arxiv} and can be explained via Eq.~(\ref{Eq:effpot}) as:
 the 't Hooft term $c_{a}\xi$ acts as a negative cubic term in the effective potential $\tilde{U}(\sigma_{x},\sigma_{y},\Phi)$,
thus a decreasing 't Hooft term will definitely accelerate the reduction of the order parameter.
And at high temperature region,
the chiral symmetry has been recovered
thus the effect of the 't Hooft term becomes negligible.
As for the deconfinement phase transition, the calculated order parameter $\tilde{\Phi}$
in Fig.~\ref{poly} displays their own similar behaviors: $\tilde{\Phi}$ increases and then deconfinement phase transition is triggered earlier as the $U_{A}(1)$ restoration is considered,
while the variation amplitudes of $\tilde{\Phi}$ are much smaller than the $\tilde{\sigma}_{x}$.

\subsection{MESON SPECTRUM AND MESON MIXING}\label{Sec:meson}
\begin{figure}[htb]
\centering
\includegraphics[width=0.45\textwidth]{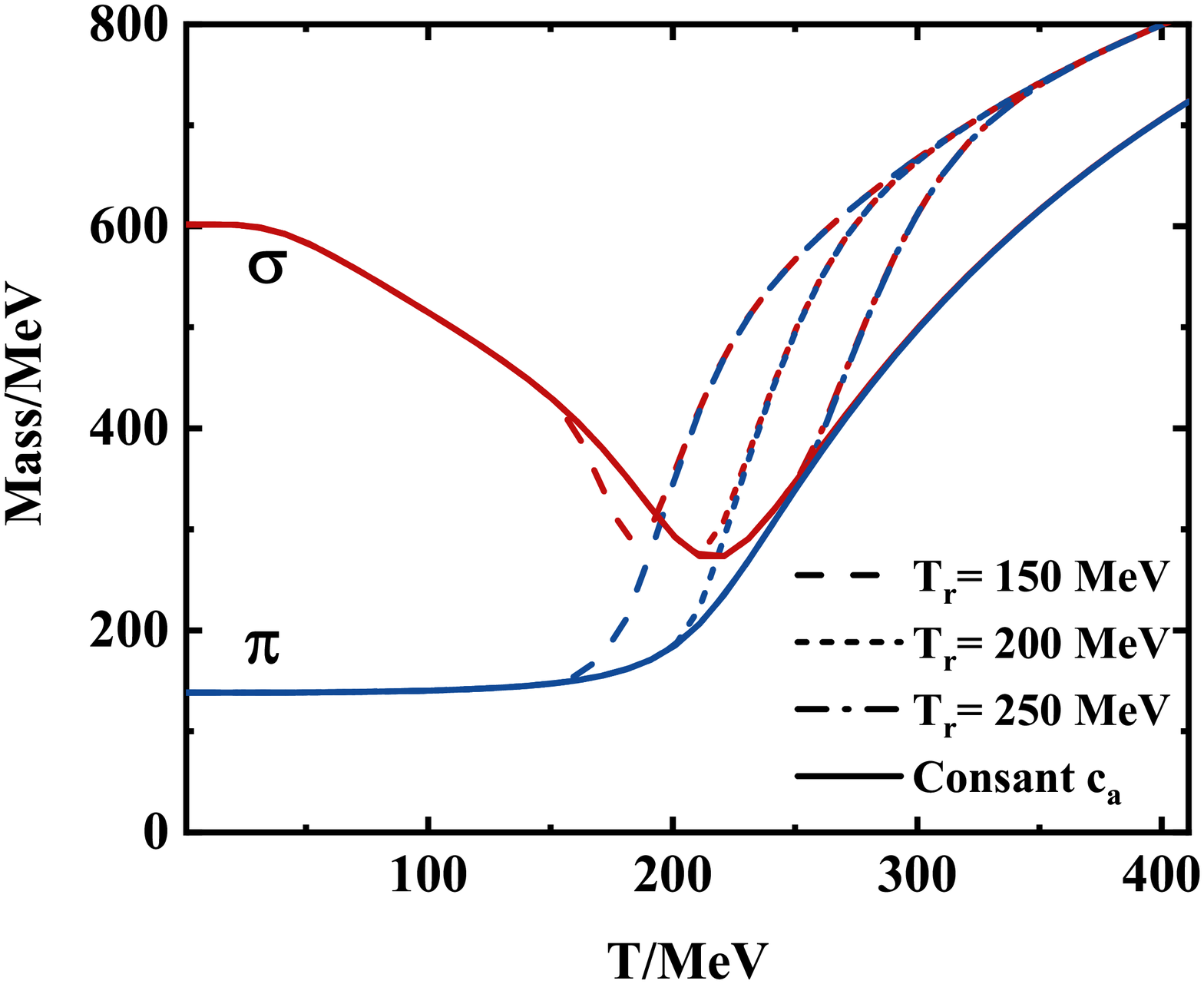}
\vspace*{-3mm}
\caption{(color online) Calculated masses of $\pi$ and $\sigma$ mesons as functions of temperature at several values of $T_{\textrm{r}}$.}\label{chiral}
\end{figure}
\begin{figure}[htb]
\centering
\includegraphics[width=0.45\textwidth]{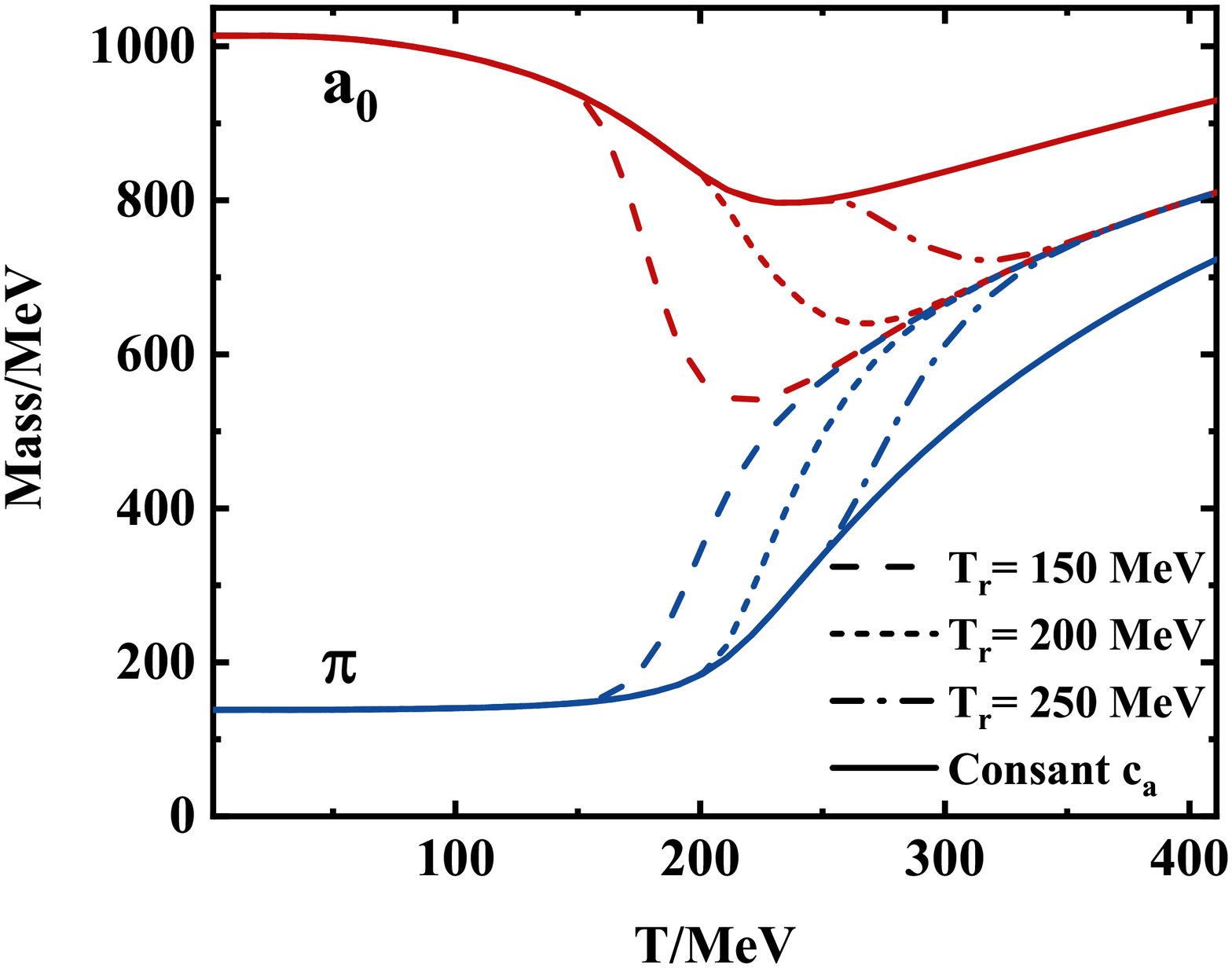}
\vspace*{-3mm}
\caption{(color online) Calculated masses of $\pi$ and $a_{0}$ mesons as functions of temperature at several values of $T_{\textrm{r}}$.}\label{axial}
\end{figure}
\begin{figure}[htb]
\centering
\includegraphics[width=0.45\textwidth]{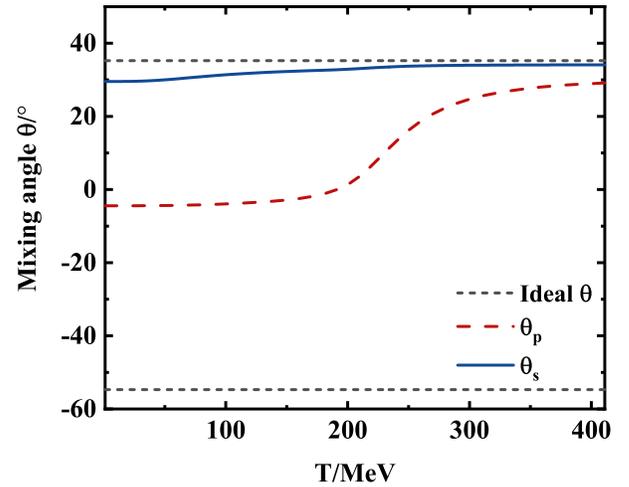}
\vspace*{-3mm}
\caption{(color online) Calculated mixing angles $\theta_{p},\theta_{s}$ as functions of temperature with a constant 't Hooft term.}\label{broken-ang}
\end{figure}
\begin{figure}[htb]
\centering
\includegraphics[width=0.45\textwidth]{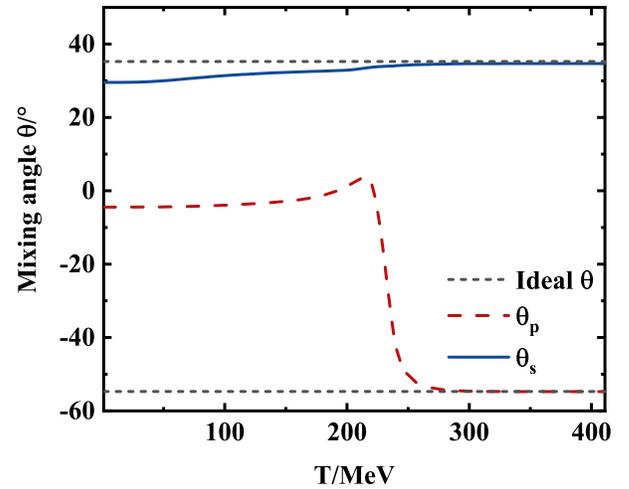}
\vspace*{-3mm}
\caption{(color online) Calculated mixing angles $\theta_{p},\theta_{s}$ as functions of temperature with a decreasing 't Hooft term, $T_{\textrm{r}}$ is chosen to be 200 MeV.}\label{restored-ang}
\end{figure}
The same effect of the $U_{A}(1)$ symmetry restoration is also manifested in meson spectrum.
It is well known that the $\sigma$ and $\pi$ mesons form a four-dimensional representation of the
$SU_{V}(2)\times SU_{A}(2)$ group and they will get degenerate when the chiral symmetry is restored. Fig.~\ref{chiral} shows evidently that the degeneration of the $\sigma$ and $\pi$ is also facilitated by the
$U_A(1)$ symmetry restoration effect, which is consistent with the behaviors of order parameters displayed in Figs.~\ref{order} and \ref{poly}.

Meson spectrum can also be implemented to explore the status of the $U_{A}(1)$ symmetry. Fig.~\ref{axial} shows the calculated spectrum of $a_{0}$ and $\pi$ meson, these two mesons have the same quantum number except for parity and they only get degenerate when axial symmetry is restored. As we can see from Fig.~\ref{axial}, when the 't Hooft term keeps constant, there is always a mass splitting $\Delta m$ between $\pi$ and $a_{0}$ which reads
\begin{eqnarray}\label{Eq:split}
\Delta m = m_{a_0}-m_{\pi}\sim c_a\tilde{\sigma}_y\, ,
\end{eqnarray}
so that axial symmetry is broken at any temperature~\cite{Pisarski:1984PRD,Bielich:2000PRL}.
After the anomaly strength $c_{a}$ grows a temperature dependence as Eq.~(\ref{Eq:anomal}),
$\Delta m$ begins to decrease.
$\pi$ will grow heavy gradually and finally get degenerate with $a_{0}$, $U_{A}(1)$ symmetry is then restored.
Comparing the results shown in Figs.~\ref{chiral} and \ref{axial},
we can notice that the degeneration temperature of ($a_{0}$, $\pi$) is always much higher than the corresponding chiral partners, the ($\sigma$, $\pi$) multiplet, which is consistent with the Lattice QCD simulations~\cite{Bhattacharya:2014PRL}.
Hence we deduce that the $U_{A}(1)$ symmetry is still broken as the chiral symmetry is restored.
The same conclusion can be drawn from a different perspective in the following subsection~\ref{Sec:pre}. %

When chiral symmetry $SU_{A}(3)$ is explicitly broken, particles belong to different representations can get mixed with each other and form the mass eigenstates.
For example, (pseudo-)scalar mesons in the meson matrix $\Sigma$ will get rotated to form physical particles, which reads
\begin{eqnarray} \label{eqn:mesonmixing}
\begin{pmatrix}
\eta\\
\eta'
\end{pmatrix}
=
\begin{pmatrix}
 \cos\,\theta_p &  -\sin\,\theta_p\\
 \sin\,\theta_p & \cos\,\theta_p
\end{pmatrix}
\begin{pmatrix}
\pi_{8} \\
\pi_{0}
\end{pmatrix}\, ,\nonumber\\
\begin{pmatrix}
f_0\\
\sigma
\end{pmatrix}
=
\begin{pmatrix}
\cos\,\theta_s & -\sin\,\theta_s\\
\sin\,\theta_s & \cos\,\theta_s
\end{pmatrix}
\begin{pmatrix}
\sigma_{8} \\
\sigma_{0}
\end{pmatrix}\, .
\end{eqnarray}
These mixing angles $\theta_{p}$ and $\theta_{s}$ are very sensitive to the status of $U_{A}(1)$ symmetry~\cite{Lenaghan:2000PRD} and can be calculated to study the effect of $U_{A}(1)$ restoration. The expressions for $\theta_{p,s}$ reads simply
\begin{eqnarray}\label{Eq:ang}
\tan(2\theta_{i})=\frac{2(m_{i}^{2})_{0,8}}{(m_{i}^{2})_{0,0}-(m_{i}^{2})_{8,8}} \ \ (i=s,p)\, ,
\end{eqnarray}
where $m_{i}^2$ with subscripts are the second derivatives of $U_0$ with respect to the corresponding fields.
We can see from Eq.~(\ref{Eq:ang}) that $\theta_{i}$ is actually an multi-value function with period $\pi/2$.
We will choose the branch cut of $\theta_{i}$ at each temperature point to ensure that the masses of mesons are continuous with the increasing temperature.

\begin{figure}[htb]
\centering
\includegraphics[width=0.45\textwidth]{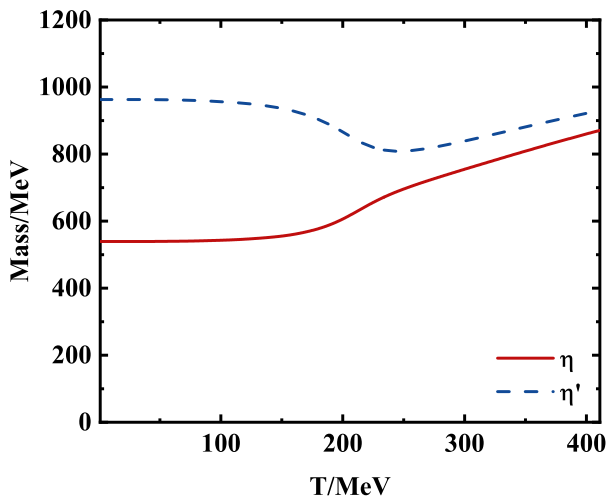}
\vspace*{-3mm}
\caption{(color online) Calculated masses of $\eta,\eta^{\prime}$ mesons with a constant 't Hooft term.}\label{broken-spec}
\end{figure}

\begin{figure}[htb]
\centering
\includegraphics[width=0.45\textwidth]{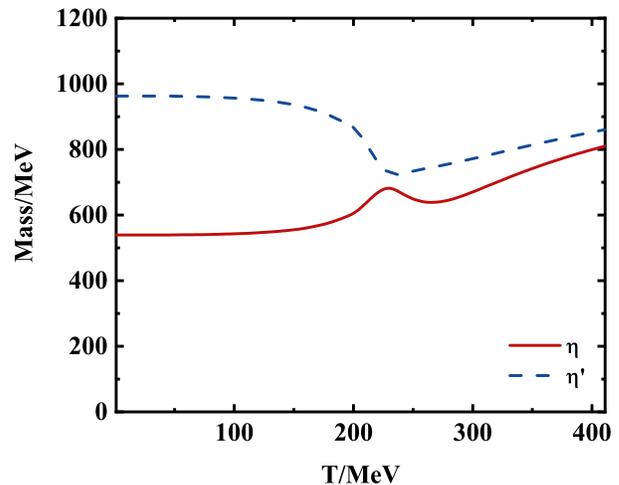}
\vspace*{-3mm}
\caption{(color online) Calculated masses of $\eta,\eta^{\prime}$ mesons with a decreasing 't Hooft term,
$T_{\textrm{r}}$ is chosen to be 200 MeV.}\label{restored-spec}
\end{figure}

The calculated mixing angles with a constant 't Hooft term are shown in Fig.~\ref{broken-ang}.
As we can see clearly, $\theta_{p}$ and $\theta_{s}$ will both increase and approach the ideal mixing angle $35^\circ$ with the ascending of temperature.
When the ideal mixing angle $35^{\circ}$ is reached, $\eta^{\prime}$ and $\sigma$ will only contain light u,d (anti)quarks while the $\eta$ and $f_{0}$ contain only s (anti)quarks according to the relation in Eq.~(\ref{eqn:mesonmixing}), which is consistent with the results in Refs.~\cite{Lenaghan:2000PRD,Schaefer:2009PRD}.
After $U_{A}(1)$ restoration effect is considered, the pseudo-scalar mixing angle $\theta_{p}$ will be changed significantly
as displayed in Fig.~\ref{restored-ang}.
It's evident that the $\theta_{p}$ will decrease to approach another ideal mixing angle $-55^{\circ}$ at high temperature, this simply means $\eta'$ will become almost strange instead at high temperature while $\eta$ will become nonstrange,
which is the same as the results in Refs.~\cite{Costa:2004PRD}. And this result is consistent with the calculated spectrum of $\eta,\eta^{\prime}$ mesons displayed in Figs.~\ref{broken-spec} and \ref{restored-spec}:  $\eta^{\prime}$ is heavier than $\eta$ at high temperature region because $\eta^{\prime}$ contains more strange quark contents than $\eta$.

Combining the calculated mixing angles and spectrum, we can see that the pseudo-scalar mesons $\eta$ and $\eta^{\prime}$
are very sensitive to the $U_{A}(1)$ symmetry.
Their mixing angles and mass spectrum can provide valuable information about the status of $U_{A}(1)$ symmetry.

\subsection{PRESSURE AND ENTROPY DENSITY}\label{Sec:pre}

\begin{figure}[htb]
\centering
\includegraphics[width=0.45\textwidth]{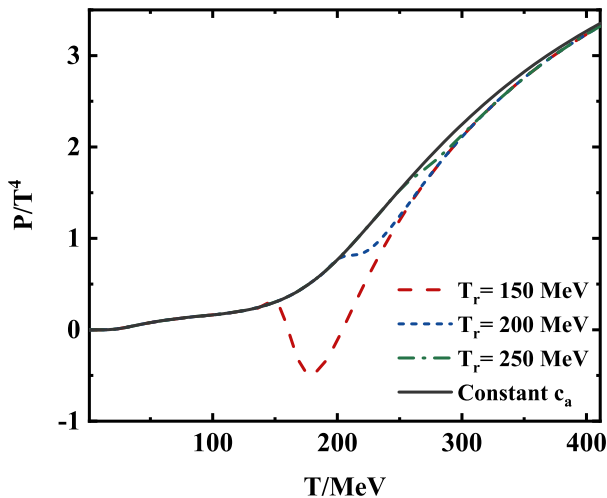}
\vspace*{-3mm}
\caption{(color online) Calculated normalized pressure as functions of temperature at several values of
$T_{\Large \textrm{r}}$.}\label{pressure}
\end{figure}

\begin{figure}[htb]
\centering
\includegraphics[width=0.45\textwidth]{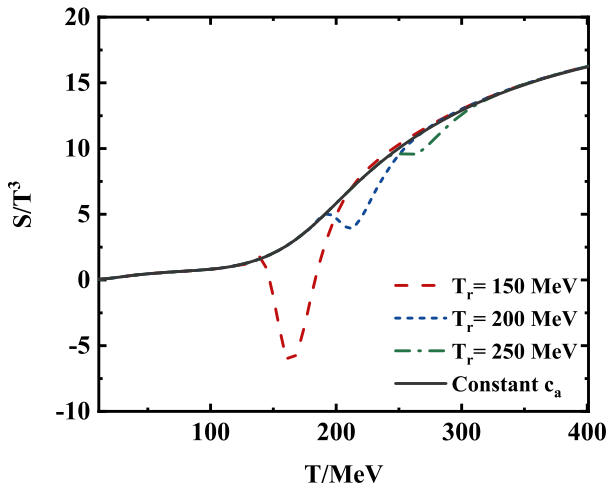}
\vspace*{-3mm}
\caption{(color online) Calculated normalized entropy density as functions of temperature at several values of
$T_{\Large \textrm{r}}$.}\label{entropy}
\end{figure}

Besides order parameters and meson\ spectrum, thermodynamical quantities such as pressure can also be calculated to manifest how does the system response to the $U_{A}(1)$ restoration.
The calculated normalized pressure is displayed in Fig.~\ref{pressure}.
The most striking aspect we obtained is that the pressure becomes negative in the chiral transition region
if the 't Hooft term begins to drop down too early.
And the $U_{A}(1)$ symmetry restoration continues to reduce the pressure at any certain temperature until the temperature is high enough, only then the pressure differences between different restoration patterns become negligible.
This unnatural behavior of the pressure is barely seen in previous Lattice QCD simulations~\cite{Bazavov:2012PRD,Borsanyi:2010JHEP}.
Since pressure is crucial for the thermodynamics of the system, other thermodynamical quantities will all be affected by such a behavior.
For example, it can be seen in Fig.~\ref{entropy} that the entropy density,
which is simply the derivative of pressure with respect to the temperature,
gets also reduced and even becomes negative if the $U_{A}(1)$ symmetry gets restored at lower temperature.
These dropping behaviors of the pressure and the entropy density can be explained via Eq.~(\ref{Eq:effpot}) as:
a dropping of the 't Hooft term lifts the bottom of the effective potential $\tilde{U}$
when the $\tilde{\sigma}_{x}$ remains sizable and then leads to a smaller pressure.
And after the chiral symmetry is restored, the $\tilde{\sigma}_{x}$ is always nearly zero
and the minimum of $\tilde{U}$ would not be affected by the 't Hooft term.

We can see from the above discussion that the contributions of the $U_{A}(1)$ symmetry breaking embodied in the instanton background of QCD are crucial for the thermodynamics of the system before the chiral phase transition
(more exactly, crossover) is completed, then even a slight drop of the $c_{a}$ will result in a unphysical pressure and entropy density at that temperature region.
After the chiral phase transition happens, the contributions of the instantons become ignorable
and then the $U_{A}(1)$ symmetry can be restored gradually without causing any unphysical results.
Thus we speculate that the $U_{A}(1)$ symmetry should be restored much later than the chiral symmetry
$SU_{A}(3)$ in order to obtain a physical pressure of the system.

\section{Summary and Remarks}\label{Sec:sum}

In this article, we investigate the effective restoration of the $U_{A}(1)$ symmetry
via the FRG approach combining with the 2+1 flavor PQM model.
The 't Hooft term $c_{a}\xi$ is parameterized as a function of temperature to imitate the $U_{A}(1)$ symmetry restoration.
Order parameters, meson spectrum and mixing angles, pressure and entropy density of the system are calculated to explore the effects of $U_{A}(1)$ restoration.

The calculated order parameters manifest that the chiral and deconfinement phase transition will be triggered
at a lower temperature if the $U_{A}(1)$ symmetry restoration happens too early,
which agrees with the predictions given in Refs.~\cite{Bielich:2000PRL,Costa:2004PRD,Rai:2018arxiv}.
The calculated meson spectrum shows that the ($a_{0},\pi$) gets degenerate later than
the ($\sigma,\pi$) multiplet and suggests a breaking of the $U_{A}(1)$ symmetry as the chiral phase transition occurs,
which is consistent with Lattice QCD simulation result~\cite{Bhattacharya:2014PRL}. Moreover, the mixing angle $\theta_p$ of $\eta,\eta'$ system is shown to be highly sensitive to the $U_{A}(1)$ restoration and can provide useful information about the status of $U_{A}(1)$ symmetry.

Besides, we provide a new insight about the $U_{A}(1)$ symmetry restoration problem:
the system will have a negative and thus unphysical pressure and entropy density if the $ U_{A}(1)$ symmetry is restored
before the chiral phase transition.
These unphysical behaviors of the pressure and entropy density can only be avoided if the $U_{A}(1)$ symmetry keeps being broken until a temperature much higher than the $T_{c}^{\chi}$ is reached.

Combining the results from meson spectrum and the thermodynamical properties,
we speculate that the $U_{A}(1)$ symmetry remains broken as the chiral symmetry $SU_{A}(3)$ gets restored.
And some underlying mechanisms are discussed.
Moreover, we would like to mention that our work only considers the physical point in the Columbia plot,
while the $U_{A}(1)$ symmetry breaking might have different fate in other region of the Columbia plot.
For example, some Lattice QCD simulations show that the $U_{A}(1)$ symmetry is restored near
the chiral phase transition in chiral limit with 2 flavor quarks~\cite{Brandt:2016JHEP,Suzuki:2018POS}.
The related investigation in FRG approach is under progress.

\begin{acknowledgments}
The work was supported by the National Natural Science Foundation of China under Contracts No.\ 11435001,
No. 11775041 , and the National Key Basic Research Program of China under Contract No.\ 2015CB856900.
\end{acknowledgments}

\end{CJK*}

\end{document}